\documentclass[aps,pre,preprint,groupedaddress,10pt]{revtex4}

\usepackage[dvips]{graphicx}

\begin{document}

\title{Calculation of the potential of mean force from nonequilibrium
measurements via maximum likelihood estimators}

\author{Riccardo Chelli, Simone Marsili, Piero Procacci}

\affiliation{Dipartimento di Chimica, Universit\`a di Firenze, Via
della Lastruccia 3, I-50019 Sesto Fiorentino, Italy}

\affiliation{European Laboratory for Non-linear Spectroscopy (LENS),
Via Nello Carrara 1, I-50019 Sesto Fiorentino, Italy}

\date{\today}

\begin{abstract}
We present an approach to the estimate of the potential of mean force
along a generic reaction coordinate based on maximum likelihood
methods and path-ensemble averages in systems driven far from
equilibrium. Following similar arguments, various free energy
estimators can be recovered, all providing comparable computational
accuracy. The method, applied to the unfolding process of the
$\alpha$-helix form of an alanine deca-peptide, gives results in good
agreement with thermodynamic integration.
\end{abstract}


\maketitle

\section{Introduction}
\label{sec:intro}

Estimate of free energy differences is useful for many applications
including protein/ligand binding affinities and drug design as well as
for theoretical perspectives. A rough classification of the plethora
of computational methods devised for determining free energy
differences can be based on the possibility of sampling a system at
equilibrium or out of equilibrium. Equilibrium approaches include
thermodynamic integration\cite{kirkwood35}, free energy
perturbation\cite{zwanzig54} and Umbrella Sampling
techniques\cite{torrie77}. Representative examples of nonequilibrium
techniques are the so-called adaptive force\cite{darve01} or
potential\cite{laio02} bias methods. The efficiency of the latter
techniques depends crucially on how fast the history-dependent force
or potential changes in time, or in other words, how far from
equilibrium the simulation is carried out. From this point of view,
adaptive bias potential methods would be more appropriately defined as
quasi-equilibrium techniques.

In the context of nonequilibrium
approaches\cite{evans93,gallavotti95}, a substantially different
scenario has been disclosed by Jarzynski\cite{jarzynski97} and
Crooks\cite{crooks98}, who introduced ``truly'' nonequilibrium methods
for determining free energy differences. In particular they proposed
two exact equations, referred here as Jarzynski equality and Crooks
nonequilibrium work theorem, relating free energy differences between
two thermodynamic states to the external work done on the system in an
ensemble of nonequilibrium paths switching between the two states. In
a recent paper Shirts {\it et al.}\cite{shirts03} have demonstrated
that the Bennett acceptance ratio\cite{bennett76} can be interpreted,
exploiting the Crooks nonequilibrium work theorem, in terms of the
maximum likelihood (ML) estimate of the free energy difference given a
set of nonequilibrium work values in the forward and reverse
directions.

One of the major shortcomings of these nonequilibrium techniques is
that free energy profile along a given reaction coordinate, {\it i.e.}
the potential of mean force (PMF), is hardly available. With the two
sets of forward and reverse nonequilibrium paths, Crooks
nonequilibrium work theorem\cite{crooks98} and ML
method\cite{shirts03} yield only the free energy differences between
the final and initial states. The Jarzynski equality, on the other
hand, can in principle be used to calculate the PMF. However it is
well-known \cite{oberhofer05,park04,hummer01,shirts05} that the
exponential average in the Jarzynski equality depends crucially on a
small fraction of realizations that transiently violate the second law
of thermodynamics. Since such ``magic'' realizations are very unlikely
to occur among a collection of fast rate realizations, it is clear
that the potential of mean force cannot be determined accurately by
the direct application of the Jarzynski equality.

In the present paper we demonstrate how to recover the PMF using ML
estimators\cite{shirts03} and path-ensemble averages in systems driven
far from equilibrium\cite{crooks00}. We test the method on the
unfolding process of the $\alpha$-helix form of an alanine
deca-peptide through steered molecular dynamics (MD) simulations.

\section{Theory}
\label{sec:theory}

\subsection{Description of the dynamical system and notation}
\label{sec:notation}

\begin{figure}[h!]
\begin{center}
\includegraphics[scale=0.45]{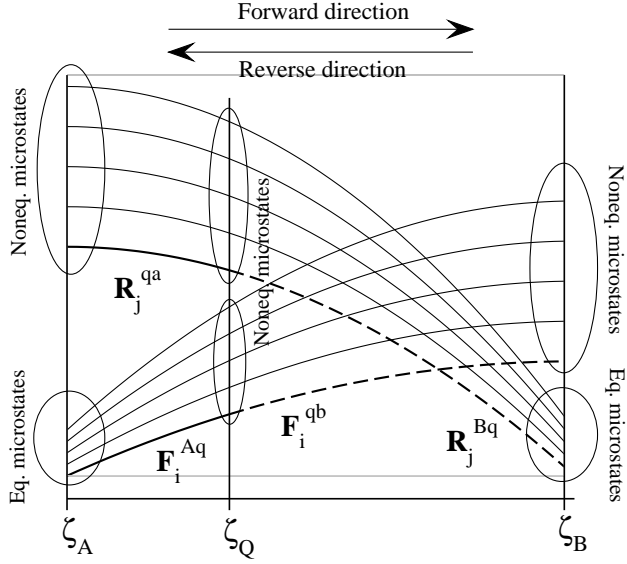}
\end{center}
\caption{Schematic representation of forward and reverse realizations
with the notation used in the text.}
\label{fig1}
\end{figure}

Let us consider a system that can switch between two states, $A$ and
$B$, characterized by different values of an arbitrary reaction
coordinate $\zeta$, namely $\zeta_A$ and $\zeta_B$. We denote with $F$
(forward) any realization during which the reaction coordinate is
forced to vary from $\zeta_A$ to $\zeta_B$ with a prescribed time
schedule. Accordingly, we denote with $R$ (reverse) any realization
that brings the reaction coordinate from $\zeta_B$ to $\zeta_A$ with
inverted time schedule. The kind of computational or experimental
technique used for producing the realizations is not relevant. The
essential requirement is that the used technique furnishes the value
of the work done on the system during the realizations. Suppose to
produce a collection of $n_F$ forward realizations and a collection of
$n_R$ reverse realizations, each realization being started from
microstates ({\it i.e.}, phase space points) sampled from an
equilibrium distribution (equilibrium microstates of $A$ for the $F$
realizations and equilibrium microstates of $B$ for the $R$
realizations). Specifically, an equilibrium microstate of, {\it e.g.},
$A$ is simply obtained by sampling the system in thermal equilibrium
with a bath, the reaction coordinate being constrained to the value
$\zeta_A$\cite{park04,procacci06}. Furthermore, we assume that all
realizations are performed at a very fast rate, which implies that
they are carried out far from equilibrium. As a consequence, the final
microstates of the $F$ and $R$ realizations will not be distributed
according to the equilibrium distribution of $B$ and $A$,
respectively. It is evident that the same holds true for the
intermediate microstates of the $F$ and $R$ realizations. For example,
the microstates characterized by a generic value $\zeta_Q$ of the
reaction coordinate obtained during a realization starting either from
$A$ ($F$ realization) or from $B$ ($R$ realization) will not be
equilibrium microstates of the state $Q$ characterized by the reaction
coordinate $\zeta_Q$. This situation is schematically represented in
Fig. \ref{fig1}. We now denote a generic $i$th $F$ realization with
${\mathbf F}_i^{Ab}$, where the superscript $Ab$ means that the
thermodynamic state corresponding to the initial microstate is $A$
(first letter) and that such microstate is taken from an equilibrium
ensemble of microstates (uppercase), while the thermodynamic state
corresponding to the final microstate is $B$ (second letter) and that
such microstate belongs to an ensemble of microstates out of
equilibrium (lowercase). Following this notation, the segments from
$\zeta_A$ to $\zeta_Q$ and from $\zeta_Q$ to $\zeta_B$ of the
${\mathbf F}_i^{Ab}$ realization are denoted as ${\mathbf F}_i^{Aq}$
and ${\mathbf F}_i^{qb}$, such that ${\mathbf F}_i^{Ab} \equiv
{\mathbf F}_i^{Aq} + {\mathbf F}_i^{qb}$. Analogously, we may write
${\mathbf R}_j^{Ba} \equiv {\mathbf R}_j^{Bq} + {\mathbf
R}_j^{qa}$. The same symbols with no specified subscripts will be used
to indicate a generic realization or a collection of realizations.
Finally, we define the free energy difference between the states $A$
and $B$ as $\Delta F_{AB} = F_B - F_A$.

\subsection{Background}

Given these two collections of nonequilibrium realizations, one may
recover $\Delta F_{AB}$ following the ML method by Shirts {\it et
al.}\cite{shirts03}. Such method is based on the maximization of the
overall likelihood of obtaining the series of measurements
(specifically the work done on the system during the $F$ and $R$
realizations) using the free energy difference as variational
parameter. In our case, the likelihood ${\mathcal L}$ of obtaining the
given work measurements can be expressed as the joint probability of
obtaining the forward measurements at the specified work values
$W[{\mathbf F}_1^{Ab}]$, $W[{\mathbf F}_2^{Ab}]$, ..., $W[{\mathbf
F}_{n_F}^{Ab}]$, times the joint probability of obtaining the reverse
measurements at the specified work values $W[{\mathbf R}_1^{Ba}]$,
$W[{\mathbf R}_2^{Ba}]$, ..., $W[{\mathbf R}_{n_R}^{Ba}]$:
\begin{equation}
\mathcal L (\Delta F_{AB} ) = \prod_{i=1}^{n_F} P\left(F ~ | ~
W[{\mathbf F}_i^{Ab}] \right) ~ \prod_{j=1}^{n_R} P \left( R ~ | ~
W[{\mathbf R}_j^{Ba}] \right),
\label{eq:likelihood1}
\end{equation}
where $W[{\mathbf F}_i^{Ab}]$ and $W[{\mathbf R}_j^{Ba}]$ are the work
performed on the system during the ${\mathbf F}_i^{Ab}$ and ${\mathbf
R}_j^{Ba}$ realizations. The best estimate of the free energy
difference $\Delta F_{AB}$ is the value that maximizes $\mathcal L
(\Delta F_{AB} )$, or equivalently its log function:
\begin{eqnarray}
\frac{\partial \ln \mathcal L (\Delta F_{AB})}{\partial \Delta F_{AB}}
 = \sum_{i=1}^{n_F} \frac{1} { 1 + \frac{n_F}{n_R} ~ {\rm e}^{ \beta (
 W[{\mathbf F}_i^{Ab}] - \Delta F_{AB} ) } } - \sum_{j=1}^{n_R}
 \frac{1}{ 1 + \frac{n_R}{n_F} ~ {\rm e}^{\beta ( W[{\mathbf
 R}_j^{Ba}] + \Delta F_{AB} ) } } = 0,
\label{eq:likelihood2}
\end{eqnarray}
where $\beta = (k_B T)^{-1}$, $k_B$ being the Boltzmann constant and
$T$ the temperature. A full derivation of above equation can be found
in Ref. \onlinecite{shirts03}. We point out that
Eq. \ref{eq:likelihood2} has been derived starting from the Crooks
nonequilibrium work theorem\cite{crooks98,crooks00}. This implies that
the time schedules of the $F$ and $R$ realizations must be related by
time reversal symmetry and that the initial microstates of the
realizations must be sampled from equilibrium distributions. Note also
that Eq. \ref{eq:likelihood2} is exactly equivalent to the Bennett
acceptance ratio method, as can be seen by comparison to Eqs. 12(a)
and 12(b) of Ref. \onlinecite{bennett76}.

\subsection{Central result}
\label{sec:central}

Suppose we want to determine the free energy difference $\Delta
F_{AQ}$ between the states $A$ and $Q$ using the $F$ and $R$
realizations introduced above. We recall that $Q$ is an intermediate
thermodynamic state between $A$ and $B$, in the sense that it is
characterized by a reaction coordinate, $\zeta_Q$, which is taken
arbitrarily from the path connecting $\zeta_A$ to $\zeta_B$, or
viceversa (see Fig. \ref{fig1}). As explained above, this free energy
difference cannot be determined simply exploiting our collections of
$F$ and $R$ realizations into Eq. \ref{eq:likelihood2}, because the
segments of the $R$ realizations generally indicated as ${\mathbf
R}^{qa}$ do not start from equilibrium microstates of the state
$Q$. However, had the $R$ realizations been started from equilibrium
microstates of $Q$, {\it i.e.}, suppose that the ${\mathbf R}^{Qa}$
realizations are available in the place of the ${\mathbf R}^{qa}$
ones, then we could apply Eq. \ref{eq:likelihood2} for the calculation
of $\Delta F_{AQ}$. In the resulting equation, which is equivalent to
Eq. \ref{eq:likelihood2} with ${\mathbf F}_i^{Ab}$, ${\mathbf
R}_j^{Ba}$ and $\Delta F_{AB}$ replaced by ${\mathbf F}_i^{Aq}$,
${\mathbf R}_j^{Qa}$ and $\Delta F_{AQ}$, respectively, the second sum
can be rearranged as follows
\begin{eqnarray}
\sum_{i=1}^{n_F} \frac{1}{ 1 + \frac{n_F}{n_R} ~ {\rm e}^{ \beta (
W[{\mathbf F}_i^{Aq}] - \Delta F_{AQ} ) } } - \int_{-\infty}^{+\infty}
\frac{ n_R \left \langle \delta \left( W[{\mathbf R}^{Qa}] - W \right)
\right \rangle} { 1 + \frac{n_R}{n_F} ~ {\rm e}^{ \beta ( W + \Delta
F_{AQ} ) } } ~ {\rm d}W = 0,
\label{eq:likelihood4}
\end{eqnarray}
where $\delta$ is the Dirac delta function and $\langle \delta (
W[{\mathbf R}^{Qa}] - W ) \rangle$ is a shorthand for $n_R^{-1}
\sum_{j=1}^{n_R} \delta(W[{\mathbf R}_j^{Qa}] - W)$. We stress again
that the initial microstates of the ${\mathbf R}^{Qa}$ realizations
are assumed to be sampled from equilibrium. Of course, since we are
dealing with nonequilibrium ${\mathbf R}^{Ba}$ realizations, work
measurements $W[{\mathbf R}^{Qa}]$ are unavailable, at least
directly. Thus, the basic problem here is to derive the unknown
quantity $\langle \delta(W[{\mathbf R}^{Qa}] - W) \rangle$ using
somehow the overall physical information contained into our ${\mathbf
R}^{Ba}$ realizations.

To this aim we take advantage of a relation due to
Crooks\cite{crooks00} that establishes a correlation between a
function of the microstate of the system determined along forward and
reverse realizations and the dissipated work done on the system during
either the forward or the reverse realizations. In particular, setting
$f[x]$ to be a function of the final microstate $x$ of a forward
realization and $f[\hat x]$ to be the same function of the initial
microstate $\hat x$ of a reverse realization, the following relation
holds:
\begin{equation}
\left \langle f[\hat x] \right \rangle_R = \left \langle f[x] ~ {\rm
e}^{ -\beta W_d } \right \rangle_F,
\label{eq:crooks4}
\end{equation}
where $W_d$ is the work dissipated during the $F$ realization. The
subscripts $F$ and $R$ indicate that the ensemble averages are
calculated on collections of forward and reverse realizations,
respectively. Therefore, since the average $\langle f[\hat x]
\rangle_R$ is computed on the initial (equilibrium) ensemble of the
reverse process, the subsequent dynamics of the system is irrelevant
and the average equals an equilibrium average of the function $f[\hat
x]$. In applying Eq. \ref{eq:crooks4} to our case, we consider the
${\mathbf R}^{Bq}$ realizations (see Fig. \ref{fig1}) as the forward
ones. This implies that the left side of Eq. \ref{eq:crooks4} refers
to an ensemble average of the equilibrium state $Q$. Moreover, for a
given microstate of the system $x_i$, corresponding to the final
microstate of the ${\mathbf R}_i^{Bq}$ realization, we set
\begin{equation}
f[x_i] = \delta ( W[{\mathbf R}_i^{qa}] - W ),
\label{eq:fxi}
\end{equation}
where $W$ is an arbitrary real number and ${\mathbf R}_i^{qa}$ is a
segment of the ${\mathbf R}_i^{Ba}$ realization. We remark that, given
a deterministic dynamical system and given a time schedule for
evolving the reaction coordinate, the quantity $\delta ( W[{\mathbf
R}_i^{qa}] - W )$ is a single value function of the microstate
$x_i$. With this provision, the general Eq. \ref{eq:crooks4} takes the
following specific form
\begin{equation}
\left \langle \delta \left( W[{\mathbf R}^{Qa}] - W \right) \right
\rangle = \left \langle \delta \left( W[{\mathbf R}^{qa}] - W \right)
~ {\rm e}^{- \beta W_d[{\mathbf R}^{Bq}] } \right \rangle,
\label{eq:crooks5}
\end{equation}
where $W_d[{\mathbf R}^{Bq}]$ is the work dissipated in the ${\mathbf
R}^{Bq}$ realizations. Since $\Delta F_{BQ}$ is unknown, $W_d[{\mathbf
R}^{Bq}]$ cannot be determined. However, upon division of
Eq. \ref{eq:crooks5} by the equality\cite{crooks00} $\left \langle
\exp \left( - \beta W_d[{\mathbf R}^{Bq}] \right) \right \rangle = 1$,
and using the definition $W_d[{\mathbf R}_i^{Bq}] = W[{\mathbf
R}_i^{Bq}] - \Delta F_{BQ}$, we obtain
\begin{equation}
\left \langle \delta \left( W[{\mathbf R}^{Qa}] - W \right) \right
\rangle = \frac{ \left \langle \delta \left( W[{\mathbf R}^{qa}] - W
\right) ~ {\rm e}^{ - \beta W[{\mathbf R}^{Bq}] } \right \rangle}
{\left \langle {\rm e}^{ - \beta W[{\mathbf R}^{Bq}] } \right
\rangle}.
\label{eq:crooks7}
\end{equation}
This equation states that the distribution of the work $W[{\mathbf
R}^{Qa}]$ done on the system in a collection of nonequilibrium
realizations switching the reaction coordinate from $\zeta_Q$ to
$\zeta_A$ and starting from equilibrium microstates can be recovered
from a set of nonequilibrium realizations switching the reaction
coordinate between the same values, but starting from nonequilibrium
microstates (realizations ${\mathbf R}^{qa}$ in
Eq. \ref{eq:crooks7}). The contribution of each work measurement
$W[{\mathbf R}^{qa}]$ to the distribution must however be weighted by
a factor depending on the work done on the system to produce the
initial nonequilibrium microstate (the factor $\exp( - \beta
W[{\mathbf R}_i^{Bq}] ) / \langle \exp(- \beta W[{\mathbf R}^{Bq}] )
\rangle$ in Eq. \ref{eq:crooks7}). We point out that, while
Eq. \ref{eq:crooks4} is valid for both stochastic and deterministic
systems\cite{crooks00}, the derivation of Eq. \ref{eq:crooks7}
provided here holds only for deterministic systems (we have indeed
introduced this assumption when defining $f[x_i]$; see
Eq. \ref{eq:fxi}). However, it has been numerically
proved\footnote{David Minh from Department of Chemistry \&
Biochemistry and Department of Pharmacology and NSF Center for
Theoretical Biological Physics, University of California San Diego,
USA; private communication.} that the relations derived in the present
article (specifically Eq. \ref{eq:likelihood12}) can also be applied
successfully to Brownian dynamical systems. Finally, substituting
Eq. \ref{eq:crooks7} into Eq. \ref{eq:likelihood4} and performing the
integral, we get
\begin{eqnarray}
\sum_{i=1}^{n_F} \frac{1}{ 1 + \frac{n_F}{n_R} ~ {\rm e}^{ \beta (
W[{\mathbf F}_i^{Aq}] - \Delta F_{AQ} ) } } - {\left \langle {\rm e}^{
- \beta W[{\mathbf R}^{Bq}] } \right \rangle}^{-1} \sum_{j=1}^{n_R}
\frac{ {\rm e}^{ - \beta W[{\mathbf R}_j^{Bq}] } } { 1 +
\frac{n_R}{n_F} ~ {\rm e}^{ \beta ( W[{\mathbf R}_j^{qa}] + \Delta
F_{AQ} ) } } = 0.
\label{eq:likelihood5}
\end{eqnarray}
The above equation is the central result of the present article. By
means of a recursive procedure, Eq. \ref{eq:likelihood5} allows us to
determine the free energy difference between the state $A$ and an
arbitrary state $Q$, and hence between any pair of states along the
reaction path. We note that in Eq. \ref{eq:likelihood5} the physical
information of both $F$ and $R$ realizations is used, albeit not at
the maximum extent. In fact, while the $R$ realizations are fully used
(note that ${\mathbf R}_j^{Bq} + {\mathbf R}_j^{qa} \equiv {\mathbf
R}_j^{Ba}$), for the $F$ realizations only the segments ${\mathbf
F}^{Aq}$ are actually employed.

An analogous and symmetrical approach aimed at using the physical
information contained into the full ${\mathbf F}^{Ab}$ realizations
and into the segment ${\mathbf R}^{Bq}$ of the ${\mathbf R}^{Ba}$
realizations allows us to recover a ML estimator to determine $\Delta
F_{QB}$:
\begin{eqnarray}
{\left \langle {\rm e}^{ - \beta W[{\mathbf F}^{Aq}] } \right
\rangle}^{-1} \sum_{i=1}^{n_F} \frac{ {\rm e}^{ - \beta W[{\mathbf
F}_i^{Aq}] } } { 1 + \frac{n_F}{n_R} ~ {\rm e}^{ \beta ( W[{\mathbf
F}_i^{qb}] - \Delta F_{QB} ) } } - \sum_{j=1}^{n_R} \frac{1}{ 1 +
\frac{n_R}{n_F} ~ {\rm e}^{ \beta ( W[{\mathbf R}_j^{Bq}] + \Delta
F_{QB} ) } } = 0.
\label{eq:likelihood6}
\end{eqnarray}
As for Eq. \ref{eq:likelihood2}, the left sides of
Eqs. \ref{eq:likelihood5} and \ref{eq:likelihood6} are strictly
increasing functions of $\Delta F_{AQ}$ and $\Delta F_{QB}$,
respectively. The limits of the left side of Eq. \ref{eq:likelihood5}
for $\Delta F_{AQ}\rightarrow \infty$ and for $\Delta
F_{AQ}\rightarrow -\infty$ are $n_F$ and $-n_R$, respectively. The
analogous limits of the left side of Eq. \ref{eq:likelihood6}, {\it
i.e.}, $\Delta F_{QB}\rightarrow \infty$ and $\Delta F_{QB}\rightarrow
-\infty$, give the same values. The monotonic behavior of the left
sides of Eqs. \ref{eq:likelihood5} and \ref{eq:likelihood6} and their
limit values guarantee the existence of one unique root. Such root
corresponds to the value of the free energy difference that furnishes
the ML estimate of the measured/calculated data. It is finally
straightforward to prove that both equations have
Eq. \ref{eq:likelihood2} as special case (set the equivalence between
the states $Q$ and $B$ in Eq. \ref{eq:likelihood5} and between the
states $Q$ and $A$ in Eq. \ref{eq:likelihood6}).

As stated above, the overall physical information available from the
$F$ and $R$ realizations is not exploited either in
Eq. \ref{eq:likelihood5} or in Eq.  \ref{eq:likelihood6}. To tackle
this fact one can however apply a ML argument to the two collections
of measurements implied by Eqs. \ref{eq:likelihood5} and
\ref{eq:likelihood6}. We first notice that Eq. \ref{eq:likelihood5}
has been derived maximizing the log function of the likelihood
${\mathcal L}(\Delta F_{AQ})$
\begin{eqnarray}
\ln {\mathcal L}( \Delta F_{AQ} ) = \sum_{i=1}^{n_F} \ln P( F ~ | ~
W[{\mathbf F}_i^{Aq}] ) + \sum_{j=1}^{n_R} \ln P^\prime(R ~ | ~
W[{\mathbf R}_j^{Qa}]).
\label{eq:likelihood7}
\end{eqnarray}
Note that in Eq. \ref{eq:likelihood7} the probability in the second
sum is primed. This means that such term must be treated with the
usual reweighting procedure (see derivation of
Eq. \ref{eq:likelihood5}) because the realizations ${\mathbf R}^{Qa}$
are unavailable (since they must start from equilibrium
microstates). The first sum of Eq. \ref{eq:likelihood7} can instead be
treated in the standard fashion, because the work measurements
$W[{\mathbf F}^{Aq}]$ are available from the full $F$ realizations.
Analogously, Eq. \ref{eq:likelihood6} has been obtained by maximizing
the log function of ${\mathcal L}(\Delta F_{QB})$
\begin{eqnarray}
\ln {\mathcal L}( \Delta F_{QB} ) = \sum_{i=1}^{n_F} \ln P^\prime( F ~
| ~ W[{\mathbf F}_i^{Qb}] ) + \sum_{j=1}^{n_R} \ln P(R ~ | ~
W[{\mathbf R}_j^{Bq}]),
\label{eq:likelihood8}
\end{eqnarray}
where the probability in the first sum is primed for the same reasons
discussed above. From a formal standpoint, Eqs. \ref{eq:likelihood7}
and \ref{eq:likelihood8} deal with two independent collections of
realizations, {\it i.e.}, ${\mathbf F}^{Aq}$ and ${\mathbf R}^{qa}$
the former equation and ${\mathbf F}^{qb}$ and ${\mathbf R}^{Bq}$ the
latter one. Therefore, noting that
\begin{equation}
\Delta F_{QB} = \Delta F_{AB} - \Delta F_{AQ}
\label{eq:deltaf}
\end{equation}
and assuming that the free energy difference $\Delta F_{AB}$ is known
(using, {\it e.g.}, the ML estimator of Eq. \ref{eq:likelihood2}), we
may express the overall likelihood of obtaining the given work
measurements in both collections as the product ${\mathcal L}(\Delta
F_{AQ}) {\mathcal L}(\Delta F_{QB})$, {\it i.e.}, a function of
$\Delta F_{AQ}$ alone (or alternatively, $\Delta F_{QB}$
alone). Maximization of the log function of such a product with
respect to $\Delta F_{AQ}$ leads to the following equation (the same
estimator would be obtained maximizing the log function of ${\mathcal
L}(\Delta F_{AQ}) {\mathcal L}(\Delta F_{QB})$ with respect to $\Delta
F_{QB}$)
\begin{equation}
\frac{\partial \ln {\mathcal L}(\Delta F_{AQ})}{\partial \Delta
  F_{AQ}} + \frac{\partial \ln {\mathcal L}(\Delta F_{QB})}{\partial
  \Delta F_{AQ}} = 0.
\label{eq:likelihood9}
\end{equation}
Since $\Delta F_{AB}$ is known, and hence independent on both $\Delta
F_{AQ}$ and $\Delta F_{QB}$, Eq. \ref{eq:deltaf} sets the condition
\begin{equation}
\frac{\partial \Delta F_{QB}}{\partial \Delta F_{AQ}} = -1.
\label{eq:likelihood10}
\end{equation}
Using Eq. \ref{eq:likelihood10} into Eq. \ref{eq:likelihood9}, leads
to the relation
\begin{equation}
\frac{\partial \ln {\mathcal L}(\Delta F_{AQ})}{\partial \Delta
  F_{AQ}} - \frac{\partial \ln {\mathcal L}(\Delta F_{QB})}{\partial \Delta
  F_{QB}} = 0.
\label{eq:likelihood11}
\end{equation}
The left and right derivatives of Eq. \ref{eq:likelihood11} are
exactly the left sides of Eqs. \ref{eq:likelihood5} and
\ref{eq:likelihood6}, respectively. Therefore, upon substitution of
Eqs. \ref{eq:likelihood5} and \ref{eq:likelihood6} into
Eq. \ref{eq:likelihood11} and using Eq. \ref{eq:deltaf} in the
resulting equation, we obtain
\begin{eqnarray}
& & \sum_{i=1}^{n_F} \frac{1}{ 1 + \frac{n_F}{n_R} ~ {\rm e}^{ \beta (
W[{\mathbf F}_i^{Aq}] - \Delta F_{AQ} ) } } - {\left \langle {\rm e}^{
- \beta W[{\mathbf R}^{Bq}] } \right \rangle}^{-1} \sum_{j=1}^{n_R}
\frac{ {\rm e}^{ - \beta W[{\mathbf R}_j^{Bq}] } } { 1 +
\frac{n_R}{n_F} ~ {\rm e}^{ \beta ( W[{\mathbf R}_j^{qa}] + \Delta
F_{AQ} ) } } - \nonumber \\
& - & {\left \langle {\rm e}^{ - \beta W[{\mathbf F}^{Aq}] } \right
  \rangle}^{-1} \sum_{k=1}^{n_F} \frac{ {\rm e}^{ - \beta W[{\mathbf
  F}_k^{Aq}] } } { 1 + \frac{n_F}{n_R} ~ {\rm e}^{ \beta ( W[{\mathbf
  F}_k^{qb}] - \Delta F_{AB} + \Delta F_{AQ} ) } } + \sum_{l=1}^{n_R}
  \frac{1}{ 1 + \frac{n_R}{n_F} ~ {\rm e}^{ \beta ( W[{\mathbf
  R}_l^{Bq}] + \Delta F_{AB} - \Delta F_{AQ} ) } } = 0.
\label{eq:likelihood12}
\end{eqnarray}
Remember that in the equation above, the quantity $\Delta F_{AB}$ must
be predetermined via Eq. \ref{eq:likelihood2}. The left side of
Eq. \ref{eq:likelihood12} is an increasing function in $\Delta F_{AQ}$
and the limits for $\Delta F_{AQ} \rightarrow \infty $ and for $\Delta
F_{AQ} \rightarrow -\infty $ have opposite signs, being $n_R + n_F$
and $-n_R -n_F$, respectively. Again, this guarantees the existence of
one unique root in the Eq. \ref{eq:likelihood12}.

If from the one hand the ML estimator of Eq. \ref{eq:likelihood12} has
the advantage of using the full physical information contained into
our sets of work measurements, on the other hand it contains the free
energy difference $\Delta F_{AB}$ that should be determined
independently. This implies that the error on the estimate of $\Delta
F_{AB}$ sums to that on $\Delta F_{AQ}$. The other derived ML
estimators, {\it i.e.} Eqs. \ref{eq:likelihood5} and
\ref{eq:likelihood6}, do not suffer of such a shortcoming. The
disadvantage is however that these ML estimators do not employ
completely the physical information of the measurements in our hands.

\section{Numerical tests: technical details}
\label{sec:details}

The ML estimators of Eqs. \ref{eq:likelihood5}, \ref{eq:likelihood6}
and \ref{eq:likelihood12} have been applied to compute the PMF for the
unfolding process of the $\alpha$-helix form of an alanine
deca-peptide (A$_{10}$) at finite temperature. Following
Ref. \onlinecite{procacci06}, we have used steered MD simulations as a
device for the numerical experiments, taking the end-to-end distance
of A$_{10}$ as reaction coordinate $\zeta$. In particular $\zeta$
corresponds to the distance between the N atom of the N-terminus
amino-acid (constrained to a fixed position) and the N atom of the
C-terminus amino-acid (constrained to move along a fixed
direction). The values of $\zeta$ in the folded and unfolded states of
A$_{10}$ are assumed\cite{procacci06} to be 15.5 and 31.5 \AA,
respectively. Moreover we have arbitrarily assumed the unfolding
process of A$_{10}$ as the forward ($F$) one. In the context of our
notation (see Sec. \ref{sec:notation}), we therefore set $\zeta_A =
15.5$ \AA~ and $\zeta_B = 31.5$ \AA. It should be noted that in
general the end-to-end distance does not determine uniquely the
configurational state of polypeptides. However, in the specific case
of A$_{10}$, the equilibrium distribution at $\zeta = \zeta_A$
corresponds to an ensemble of microstates tightly peaked around the
$\alpha$-helix form, as for this end-to-end distance alternative
structures are virtually impossible. The same holds true for the state
corresponding to $\zeta = \zeta_B$, which basically represents an
almost fully elongated configuration of the peptide. This implies that
these two thermodynamic states can be effectively sampled using
relatively few microstates obtained from equilibrium MD simulations at
the given $\zeta$ values.  The starting microstates for the $F$ and
$R$ realizations have been randomly picked (every 5 ps) from two
standard MD simulations constraining $\zeta$ to $\zeta_A$ and to
$\zeta_B$, respectively, by means of a stiff harmonic potential (force
constant equal to 800 kcal mol$^{-1}$ \AA$^{-2}$). In both equilibrium
MD simulations and in the subsequent steered MD simulations, constant
temperature (300 K) has been enforced using a Nos\'e-Hoover
thermostat\cite{hoover85,hoover86}. Force field has been taken from
Ref. \onlinecite{mackerrell98}. Given the limited size of the sample,
no cutoff radius has been imposed to the atomic pair interactions and
no periodic boundary conditions have been applied.

For each type of process, $F$ and $R$, we have generated 10$^4$
realizations guiding $\zeta$ from $\zeta_A$ to $\zeta_B$ ($F$
realizations) or from $\zeta_B$ to $\zeta_A$ ($R$ realizations) using
a harmonic potential dependent on time:
\begin{equation}
V(t) = \frac{k}{2} [ \zeta - \lambda(t) ]^2,
\label{eq:steering_potential}
\end{equation}
where the force constant $k$ is reported above. The time-dependence of
the steering parameter $\lambda(t)$ determines the pulling speed of
the nonequilibrium realizations and in general their time schedule. In
our case, $\lambda(t)$ varies linearly with the time, {\it i.e.}
$\lambda(t) = \zeta_A + \dot \lambda t$ for the $F$ realizations and
$\lambda(t) = \zeta_B - \dot \lambda t$ for the $R$ ones.  The work
measurement at a given instant $t$ of a realization is calculated
integrating the partial derivative of $V(t)$ with respect to time from
the time zero to the time $t$. Six series of $R/F$ work measurements
differing only in the pulling speed $\dot \lambda$ have been performed
($\dot \lambda = 80$, 160, 320, 533, 800, and 1600 \AA~ ns$^{-1}$).

\section{Numerical tests: results}
\label{sec:results}

\begin{figure}[h!]
\begin{center}
\includegraphics[scale=0.5]{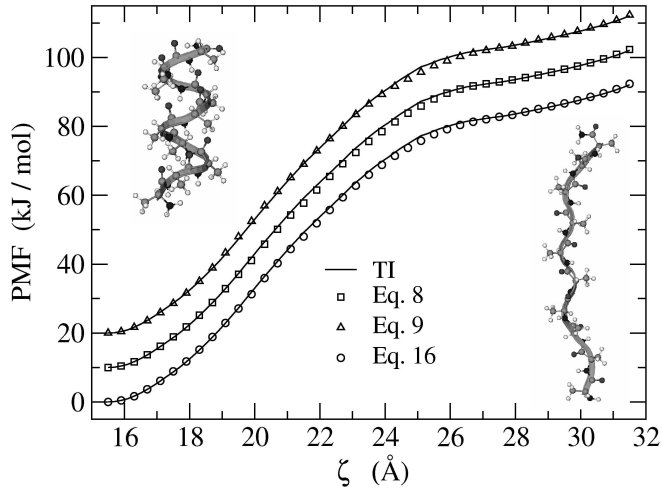}
\end{center}
\caption{PMF of A$_{10}$ as a function of the reaction coordinate
$\zeta$ (end-to-end distance). Squares: Eq. \ref{eq:likelihood5};
triangles: Eq. \ref{eq:likelihood6}; circles:
Eq. \ref{eq:likelihood12}; solid lines: thermodynamic integration
(TI). The PMF profiles from ML estimators are calculated using $F$ and
$R$ realizations performed with pulling speed of 80 \AA~
ns$^{-1}$. For the sake of clarity the curves are up-shifted.}
\label{fig2}
\end{figure}

\begin{figure}[ht]
\begin{center}
\includegraphics[scale=0.5]{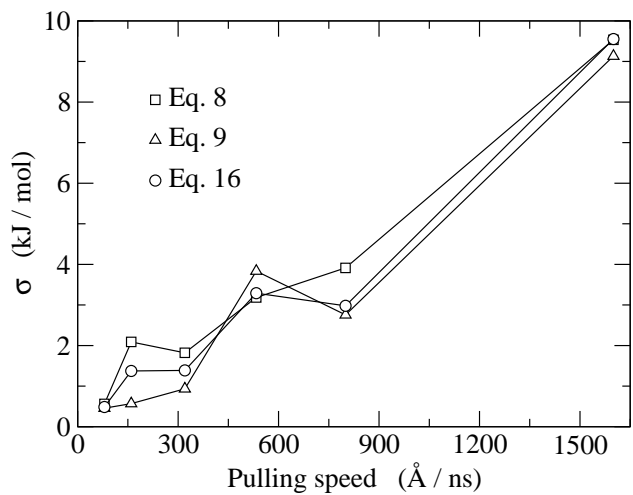}
\end{center}
\caption{$\sigma$ value (Eq. \ref{eq:sigma}) as a function of the
pulling speed for various ML estimators. Squares:
Eq. \ref{eq:likelihood5}; triangles: Eq. \ref{eq:likelihood6};
circles: Eq. \ref{eq:likelihood12}. The lines are drawn as a guide for
eyes.}
\label{fig3}
\end{figure}

In Fig. \ref{fig2} we report a comparison between the PMF calculated
using the ML estimators of Eqs. \ref{eq:likelihood5},
\ref{eq:likelihood6} and \ref{eq:likelihood12} and the exact PMF
recovered through thermodynamic integration\cite{kirkwood35}.
To show the correctness of the estimators numerically, in the figure
we have drawn the PMF profiles obtained using the slowest pulling
speed, {\it i.e.} 80 \AA~ ns$^{-1}$. It is noticeable the all ML
estimators provide an almost perfect agreement with thermodynamic
integration. The relevance of this result is enforced to the light of
early PMF calculations\cite{park04} on the unfolding process of
A$_{10}$. Free energy estimators, such as Jarzynski equality and
second order cumulant expansion\cite{park04}, using only a slightly
faster pulling speed (100 \AA~ ns$^{-1}$), give a much worse accuracy
than the methods proposed here. This can be more strictly verified
comparing Fig. 5a of Ref. \onlinecite{park04} with the PMF estimates
determined using Eqs. \ref{eq:likelihood5}, \ref{eq:likelihood6} and
\ref{eq:likelihood12} with pulling speed of 160 \AA~ ns$^{-1}$ (see
supplementary material). A more comprehensive view of the performances
of the ML estimators of Eqs. \ref{eq:likelihood5},
\ref{eq:likelihood6} and \ref{eq:likelihood12} is gained by the root
mean square deviation of the estimated PMF curves from the exact one:
\begin{equation}
\sigma = \left[ \frac{1}{N} \sum_{i=1}^N \left( {\mathcal F}_{\rm
ML}(\zeta_i) - {\mathcal F}_{\rm TI}(\zeta_i) \right)^2
\right]^{\frac{1}{2}},
\label{eq:sigma}
\end{equation}
where ${\mathcal F}_{\rm ML}(\zeta_i)$ is the value of the PMF at
$\zeta = \zeta_i$ calculated using one of our ML estimators and
${\mathcal F}_{\rm TI}(\zeta_i)$ is the corresponding exact value
determined by thermodynamic integration. In our calculations, the
reaction coordinate is defined in steps of 0.4 \AA, {\it i.e.}
$\zeta_1 \equiv \zeta_A = 15.5$, $\zeta_2 = 15.9$, $\zeta_3 = 20.3$,
$\cdots$, $\zeta_N \equiv \zeta_B = 31.5$ \AA, where $N = 41$. The
value of $\sigma$ for the various approaches has been calculated after
determining the additive constant of ${\mathcal F}_{\rm ML}(\zeta)$
via a least squares fitting to ${\mathcal F}_{\rm TI}(\zeta)$. The
value of $\sigma$ obtained from the considered ML estimators for
different pulling speeds is reported in Fig. \ref{fig3}. 
The worsening of the accuracy of the ML estimators by increasing the
pulling speed of the realizations is expected on the basis of
statistical reasons. The remarkable result is that all ML estimators
have comparable accuracy independing on the pulling speed. Moreover,
Eq. \ref{eq:likelihood6} gives the best accuracy for all pulling
speeds except for 533 \AA~ ns$^{-1}$, while Eq. \ref{eq:likelihood12}
gives systematically an accuracy which is in between those obtained
from Eqs. \ref{eq:likelihood5} and \ref{eq:likelihood6}. These facts
suggest that the formal asymmetry of the ML estimators of
Eqs. \ref{eq:likelihood5} and \ref{eq:likelihood6} (see discussion in
Sec. \ref{sec:central}) in comparison to the formal symmetry of
Eq. \ref{eq:likelihood12} may be relevant in the choice of the most
accurate approach. In fact, it is known\cite{shirts05} that, for a
given reaction path of a system, the use of a set of forward
realizations in the framework of exponential averages for determining
the free energy difference between two states may give different
variance and bias with respect to the same estimate performed using
the reverse realizations. We do not exclude that this fact might be
related to our observations of Fig. \ref{fig3} discussed above. In
such a case the ML estimator of Eq. \ref{eq:likelihood12} would be the
more appropriate without prior knowledge on the system.

\section{Conclusions}
\label{sec:conclusions}

We have presented a method for determining the PMF along a given
reaction coordinate which is based on ML methods and path-ensemble
averages in systems driven far from equilibrium. The method has been
applied to computer experiments on the unfolding process of the
$\alpha$-helix form of an alanine deca-peptide using nonequilibrium
realizations with various pulling speeds. The estimated PMF is in fair
agreement with thermodynamic integration. A formula for the variance
of PMF estimates generated using this method is still unavailable,
though its derivation appears straightforward following the guidelines
reported in the present article and in the work by Shirts {\it et
al.}\cite{shirts03}. We plan to report on this issue in a forthcoming
contribution.

\begin{acknowledgments}
We thank David Minh (Department of Chemistry \& Biochemistry and
Department of Pharmacology and NSF Center for Theoretical Biological
Physics, University of California San Diego, USA) for providing
insights about the generality of the ML estimators and for suggestions
on how improve some parts of the manuscript. This work was supported
by the European Union (Grant No. RII3-CT-2003-506350).
\end{acknowledgments}


\end{document}